# Direct electrocaloric, structural, dielectric, and electric properties of lead-free ferroelectric material $Ba_{0.9}Sr_{0.1}Ti_{1-x}Sn_xO_3$ synthesized by semi-wet method

H. Zaitouni[1], L. Hajji[1,*], D. Mezzane[1], E. Choukri[1], A. Alimoussa[1], S. Ben Moumen[1], B. Rožič[2],

M. El Marssi[3], Z. Kutnjak[2]

[1] Laboratory of Condensed Matter and Nanostructures (LMCN), Cadi-Ayyad University, Faculty of Sciences and Technology, Department of Applied Physics, Marrakech, Morocco.

[2] Laboratory for calorimetry and dielectric spectroscopy, Condensed Matter Physics Department, Jozef Stefan Institute, Ljubljana, Slovenia.

[3] Laboratoire de physique de la matière Condensée (LPMC), Université de Picardie Jules Verne, 33 rue Saint-Leu, 80039 Amiens Cédex, France.

**Abstract**

By using the semi-wet synthesis method, lead-free ferroelectric materials $Ba_{0.9}Sr_{0.1}Ti_{1-x}Sn_xO_3$ with x = 0, 0.02, 0.05, and 0.10 (abbreviated as BSTS) were prepared and their structural, electric and electrocaloric properties were investigated. The X-ray diffraction (XRD) patterns show that the samples calcined at 950°C have well crystallized into perovskite structure suggesting the substitution of $Ti^{4+}$ by $Sn^{4+}$ in BST lattice. With increasing content of Sn, the enhancement of the dielectric permittivity was observed for ($0 \leq x \leq 0.05$) and the ferroelectric transition temperature ($T_C$) was found to shift towards the room temperature ($T_C = 20°C$ for x = 0.10). Direct measurements of the electrocaloric effect (ECE) were performed on all samples by using the high-resolution calorimeter. It is found that $Ba_{0.9}Sr_{0.1}Ti_{0.95}Sn_{0.05}O_3$ exhibits a high ECE temperature change of $\Delta T_{EC} = 0.188$ K at an applied electric field of only 7 kV/cm. Impedance spectrum analysis of all the samples performed in the temperature range of 300-360°C reveals the existence of two relaxation contributions related to the grain and grain boundaries that are well separated in frequency. Activation energies of conduction and relaxation processes were deduced for both contributions in order to determine the conduction mechanism of the studied compositions.

* Corresponding author: Pr. Lahoucine Hajji

Phone: (+212) 6 68 88 55 54,  E-mail: l.hajji@uca.ma; hajji1966@gmail.com

## 1. Introduction

The electrocaloric effect (ECE) is the temperature change of a dielectric material induced by a bias electric field application [1, 2]. It can be an adiabatic change if the electric field is applied or removed under adiabatic conditions. The phenomena of ECE is of the great importance for development of new generation of cooling and heating devices that can be more environmentally friendly. Over the past decades, ECE was studied in various dielectric materials, but it was less attractive due to its small magnitude. Recently, the giant ECE response was observed in some organic and inorganic bulk materials, and also in PZT thin films with $\Delta T = 12$ K at 222°C [3–8] via the indirect electrocaloric measurements that are based on measurements of the electric polarization versus temperature and electric field P(T,E). Lately, it was shown that a high electrocaloric response can be found also in lead-free materials such as the ferroelectric material $BaTiO_3$ reported in Moya *et al*. [9], which exhibit an adiabatic temperature change of ($\Delta T = 0.87$ K) around 397 K under an electric field of 4 kV/cm.

In addition to the high electrocaloric response, many lead-free ferroelectric materials also exhibit interesting piezoelectric, pyroelectric and energy storage properties [10–13]. Barium Strontium Titanate with general formula $Ba_{1-x}Sr_xTiO_3$ (BST) is among the most widely studied ferroelectric materials. It is extensively used in several applications such as multilayer ceramic capacitors, tunable microwave devices and also as insulators in dynamic random-access memory (DRAM) [14–19]. BST is a ferroelectric material that exhibits high dielectric constant ($\varepsilon'$) and has a tunable Curie temperature ($T_c$) by modifying the mole fraction of $Ba^{2+}/Sr^{2+}$. In particular, the composition $Ba_{0.9}Sr_{0.1}TiO_3$ studied in our earlier work [20] shows a ferroelectric properties with a phase transition at $T_C = 95$°C.

It is known that the dielectric, structural and electrocaloric properties of BST can be controlled not only by the Ba/Sr molar ratio, but also through the introduction of aliovalent or isovalent cations on the B-sites of the perovskite lattice. It was shown in literature that the substitution of $Ti^{4+}$ by $Sn^{4+}$ in the $BaTiO_3$ system can have a significant effect on the electrocaloric response, dielectric, structural properties as well as it can move the Curie temperature to lower temperatures [21–23]. It is observed that $BaTi_{1-x}Sn_xO_3$ (BTS) system has a sharp phase transition up to x = 0.05, which becomes broader and diffused with increasing Sn concentration [21].

In contrast to BTS, only few works have investigated the effect of the partial substitution of $Ti^{4+}$ by $Sn^{4+}$ in the BST matrix [24–26]. In the present work, the impact of Sn on the electrocaloric, structural, dielectric and electric conductivity properties of $Ba_{0.9}Sr_{0.1}TiO_3$ was investigated.

$Ba_{0.9}Sr_{0.1}Ti_{1-x}Sn_xO_3$ with x = 0, 0.02, 0.05 and 0.10 ceramics were synthesized by a semi-wet route at a relatively low temperature as compared to the solid state method. The semi-wet route is a modified chemical method that is usually used to prepare the $ACu_3Ti_4O_{12}$ (A = Ca, Sr, Ba, $Y_{2/3}$, etc.) ceramics [27, 28]. This synthesis method consists of mixing nitrate, chloride or acetate solutions of constituents with solid $TiO_2$ powder, which is of low cost and insoluble in solvents. In contrast to the sol-gel method in which solutions of constituents are mixed with titanium isopropoxide that is more expensive. The BSTS samples were characterized by the X-ray diffractometry (XRD), scanning electron microscopy (SEM), high resolution calorimetry, and impedance spectroscopy in order to determine their structural, electrocaloric, dielectric, and electric conductivity properties.

## 2. Experimental procedure

Polycrystalline $Ba_{0.9}Sr_{0.1}Ti_{1-x}Sn_xO_3$ ceramics with x = 0, 0.02, 0.05 and 0.10, denoted as BSTS0, BSTS2, BSTS5 and BSTS10, respectively, were synthesized by the semi wet route. In this method, chemicals Barium nitrate $Ba(NO_3)_2$, strontium nitrate $Sr(NO_3)_2$, tin chloride $SnCl_2.2H_2O$, titanium oxide $TiO_2$ and citric acid were taken in stoichiometric ratios. Solutions of $Ba(NO_3)_2$, $Sr(NO_3)_2$ and $SnCl_2.2H_2O$ were prepared by using distilled water. The solutions were mixed together in a beaker and the stoichiometric amount of solid $TiO_2$ and citric acid equivalent to metal ions were also added. The obtained solution was heated on a hot plate with magnetic stirrer at a moderate temperature between 70°C-80°C and then the mixture was dried at 100°C-120°C in the hot air oven. The dried powders were calcined for 12h at 950°C. After the addition of 5 wt% of polyvinyl alcohol (PVA) to the calcined powders as a binder, the powders were pressed into disks having 1-2 mm thickness and 13 mm dimeter by using the uniaxial hydraulic press. The pellets were burned out at the temperature of 700°C to remove PVA and then sintered at 1350°C in air for 7h.

For direct electrocaloric measurements a high resolution calorimeter was utilized allowing high resolution measurements of the sample-temperature variation due to the ECE induced by a change in the applied bias electric field [29, 30]. The thickness of bulk BSTS ceramics was around ~1 mm. The samples were covered with silver electrodes and the temperature was measured using a small bead thermistor as described in the reference [31].

The dielectric measurements were carried out by using an impedance meter HP 4284A connected to the computer, in the frequency range of 20 Hz-1 MHz. The samples were characterized by XRD using the Rikagu SmartLab diffractometer with $CuK_\alpha$ ($\lambda K\alpha1$ = 1.540593 Å and $\lambda K\alpha2$ = 1.544414 Å). The lattice parameters were calculated and refined by the FullProf program. The ceramic bulk density was measured by using Archimedes principles and the grain morphology was determined by using Scanning electron microscope TESCAN VAGA3.

## 3. Results and discussion

### 3.1 Structural properties

The crystalline structure of BSTS ceramics with x = 0, 0.02, 0.05, and 0.10 was investigated by XRD. Fig. 1(a) shows room temperature XRD patterns of BSTS samples calcined at 950°C. All the samples show single perovskite phase without any traces of impurities as previously reported in several references [20, 25], suggesting the substitution of $Ti^{4+}$ by $Sn^{4+}$ in the BST lattice. With increasing Sn content, the diffraction peaks shift towards lower diffraction angles (2θ) (Fig. 1b), which is due to the partial replacement of the $Ti^{4+}$ ions with small radius (r = 0.605 Å) by the $Sn^{4+}$ ions with larger radius (r = 0.690 Å). The {200} peaks around 2θ = 45° are generally used to distinguish Rhombohedral (R), Orthorhombic (O), Tetragonal (T) and Cubic (C) phases. From the Fig. 1(a) it can be observed that the BSTS system exhibits at room temperature the T phase (P4mm space group) for x = 0 and 0.02. For x = 0.05 coexistence of the T (P4mm) and O (Amm2) phases is observed, while the composition x = 0.10 exhibits the C phase (Pm-3m).

**(Insert Fig. 1(a, b), here)**

The evolution of the lattice parameters and the unit-cell volume are shown in Fig. 2 and the values are reported in Table 1.

**(Insert Fig. 2, here)**

**(Insert Table 1, here)**

SEM images of BSTS ceramics for different Sn-content are shown in Fig. 3 (a-d). The samples with x = 0, 0.02 and 0.05 show regularly shaped grains with clear grain boundaries and an average grain size gradually increasing from ~2 µm to ~3.5 µm. However, for higher Sn doping (x = 0.10), the microstructure exhibits an inhomogeneous grain distribution with large and small grains. The relative density ($\rho_r$), defined as the ratio between the ceramic bulk density ($\rho_b$) and the theoretical density ($\rho_{th}$) of the studied samples was calculated as follow,

$$\rho_r = \frac{\rho_b}{\rho_{th}} \times 100. \quad (1)$$

The theoretical density $\rho_{th}$ was determined from X-ray measurements. The values of ceramic densities of BSTS samples are listed in Table 2. It is clearly seen that the relative density of the obtained ceramics is in the range of 94-96% which confirms the high densification of the samples. Since the BSTS samples exhibit nearly the same relative density, then the increase in grain size is mainly dependent on the substitution of $Ti^{4+}$ by $Sn^{4+}$.

**(Insert Fig. 3 (a-d), here)**

**(Insert Table 2, here)**

### 3.2 Dielectric measurements

The temperature profiles of the dielectric constant (ε') and the dielectric loss (tgδ) of all BSTS (x = 0, 0.02, 0.05 and 0.10) at several frequencies are shown in Fig. 4(a-d). The ε' increases with increasing temperature to reach a maximum value of the permittivity (ε'$_{max}$) at T$_C$ which corresponds to the transition temperature from the ferroelectric to the paraelectric phase.

A comparative study of the dielectric constant as a function of temperature for all the compositions at 1 KHz is represented in Fig. 5. The T$_C$ is shifted towards lower temperatures with increasing Sn-content. This behavior is in a good agreement with literature [21, 25]. It should be noted that with increasing Sn doping an initial increase in the maximum value of the dielectric constant is observed followed by a decrease and broadening of the dielectric peak. These results are in good agreement with SEM results showing also an increase in the grain size with a maximum value observed for x

= 0.05 that is followed by a drop in grain size. Moreover, the analysis of the dielectric measurements for the composition x = 0.05 reveals the existence of another anomaly below the Tetragonal-Cubic transition (T-C) corresponding to the Orthorhombic-Tetragonal transition (O-T) at around $T_{O-T} \approx 28°C$, which is in good agreement with the XRD suggesting the coexistence of O-T phases at room temperature (RT).

Because of the partial replacement of smaller $Ti^{4+}$ ions by larger $Sn^{4+}$ ions in B sites of the perovskite structure and because in the $Ba(Ti_{1-x}Sn_x)O_3$ the overlap between Sn and oxygen atoms is weaker than between Ti and oxygen atoms [32], the weakening of the bonding force B-O is taking place with introducing more Sn in the structure. In this case, the B site ion can reach its equilibrium position only at lower temperature, which explains the decrease of the $T_C$.

**(Insert Fig. 4 (a-d), here)**

**(Insert Fig. 5, here)**

Inset in Fig. 5 shows the evolution of RT permittivity ($\varepsilon'_{RT}$) as a function of the composition x. An increase in the RT permittivity values with increasing Sn-content from 2516 for BSTS0 to 7508 for BSTS10 is observed. The highest $\varepsilon'_{RT}$ value is found in the composition BSTS10, because of its $T_C$ ($T_C$ = 293 K) near the RT.

It is well known that in the paraelectric phase, the evolution of dielectric permittivity as a function of temperature is well described by Curie-Weiss law given by the following equation.

$$\varepsilon' = \frac{C}{T-T_0}, \qquad (2)$$

where C and $T_0$ are the Curie-Weiss constant and Curie temperature, respectively. Fig. 6 shows the variation of the 1/ε' with temperature for the different compositions of BSTS at 1 KHz. It can be observed that the experimental data are well fitted using the Eq. (2). The fitted parameters are listed in Table 3. It is evident from the obtained results that BSTS with x≤0.05 exhibits a transition of the first order ($T_0<T_C$), while it changes to the second order for x = 0.10 ($T_0 \approx T_C$).

**(Insert Fig. 6, here)**

Furthermore, to correlate the SEM with dielectric results, the diffuseness of the ferroelectric-paraelectric (F-P) phase transition for all the samples was studied by using the Santos-Eiras equation [33]

$$\varepsilon' = \frac{\varepsilon'_m}{1+\left(\frac{T-T_m}{\Delta}\right)^\gamma}, \qquad (3)$$

where $\varepsilon'_m$ is the maximum dielectric constant at the transition temperature $T_m$, $\gamma$ indicates the character of the F-P phase transition and $\Delta$ is the peak broadening which reflects the degree of diffuseness of the F-P phase transition. The temperature profiles of the dielectric constant were fitted by Santos-Eiras equation (Eq. 3) for different compositions of BSTS (Fig. 5). Good agreement between experimental data and the fit was observed.

The fitted parameters were listed in Table 3. The obtained $\gamma$ value is found to be around 1 for pure BST (BSTS0), which indicates that BST is a normal ferroelectric [20]. In contrast, the $\gamma$ value is found to be between 1 and 2 when increasing Sn-content thus indicating an incomplete ferroelectric phase transition. The values of the peak broadening $\Delta$ for BSTS showed a little increase with the different compositions x, in particular for the composition x = 0.10, where $\Delta$ is found to be higher than in the other compositions indicating that the incorporation of Sn increases the diffuseness degree of BSTS sample. According to literature, the diffuseness is generally related to the variation in local composition leading to different microregions with various local Curie temperatures for the F-P transition [34, 35]. However, it should be noted that no pronounced relaxor behavior is observed for the different compositions *x*. The same behavior is observed for BaTi$_{1-x}$Sn$_x$O3 ceramics, (x = 0.10-0.20) reported in Shvartsman *et al* [36].

**(Insert Table 3, here)**

### 3.3 Direct electrocaloric measurements

The direct electrocaloric response (electrocaloric temperature change $\Delta T_{EC}$) versus temperature investigated only in three studied samples BSTS0, BSTS2 and BSTS5 under an applied electric field of 7 kV/cm is shown in Fig. 7. The EC response shows a maximum near the F-P transition as determined by the dielectric measurements. The results show that BSTS5 ceramic exhibits the highest EC response in comparison with the two others ceramics (BSTS0 and BSTS2). The electrocaloric responsivity ($\xi = \Delta T_{EC}/\Delta E$) values calculated for the investigated compositions are summarized in Table 4 in comparison with previously reported values for other lead-free materials. It is found that the EC responsivity from this work is comparable and/or even higher than that reported for other lead-free ferroelectric materials. Furthermore, BSTS5 which exhibits the highest EC responsivity ($\xi = 0.027$ K cm/kV) at low applied electric field and shows a transition temperature ($T_C$) near the room temperature can represent a potential candidate for coolant material in novel electrocaloric refrigerators.

**(Insert Fig. 7, here)**

**(Insert Table 4, here)**

### 3.4 Impedance spectroscopy

In order to get better understanding on the electric characteristics of Sn doped BST ceramics, the impedance spectroscopy was performed. Fig. 8(a-d) shows the Nyquist plots ($-Z_i$ vs. $Z_r$) of BSTS samples measured in the temperature range of 300-360°C. All samples exhibit two semicircles, which can be attributed to the grain and grain boundary contributions at high and lower frequencies, respectively. The experimental data were well fitted (by using the Z-view software) with an electric equivalent circuit that is consisting of two parallel combinations of the resistance and a constant phase element (R//CPE) connected in series of all BSTS compositions (insets of Fig. 8(a-d)). The resistances of both grain ($R_g$) and grain boundaries ($R_{gb}$) for the studied samples are plotted as a function of inverse of temperature in Fig. 9 and in the inset to Fig. 9, respectively. We observe that both $R_g$ and $R_{gb}$ are found to decrease with increasing temperature for all the samples, which indicate the semiconducting behavior of BSTS. In order to investigate the activation energy process ($E_a^R$) for $R_g$ and $R_{gb}$, we use the following equation that satisfies the Arrhenius law,

$$R = R_0 \exp(E_a^R/k_B T) \,. \tag{4}$$

Here $R_0$ is the pre-exponential factor and $k_B$ is the Boltzmann constant. The values of the activation energies for both contributions for all BSTS samples were summarized in Table 5. It is known that in the perovskite structure, oxygen vacancies represent one type of mobile charges. At higher temperatures the singly and/or doubly ionized oxygen vacancies are induced with activation energies ranging from 0.3-0.5 eV and 0.6-1.2 eV, respectively [40, 41]. In the studied samples, it was observed that the activation energy for grain and grain boundary resistances was found to decrease with increasing Sn-content. The values of activation energy are ranging between ~0.6 eV and ~1.14 eV for the compositions x = 0, 0.05 and 0.10, which is coherent with doubly ionized oxygen vacancies. However, the higher activation energy found for both $R_g$ and $R_{gb}$ (~1.6 eV) in the composition (x = 0.02) can be attributed to the conduction process associate with barium vacancies [42].

**(Insert Fig. 8 (a-d), here)**

**(Insert Fig. 9, here)**

The relaxation times for the grain and grain boundaries of BSTS samples were also deduced from the fitting parameters (Z-View software). The formula $(\tau_i = (R_i Q_i)^{\frac{1}{n_i}})$ is employed to calculate the relaxation time for both contributions. Fig. 10 displays the linear behavior of $\ln\tau$ vs. $1/T$ for all the samples, which obey the Arrhenius law given by the expression

$$\tau = \tau_0 \exp(E_{relax}/k_B T), \qquad (5)$$

where $\tau_0$ is the pre-exponential term, $k_B$ is the Boltzmann constant and $E_{relax}$ is the activation energy for the relaxation process. Table 5 shows that the activation energies for relaxation of the grain and grain boundaries are ranging from ~0.6 eV to ~1.3 eV for the compositions x = 0, 0.05 and 0.10. For BSTS x = 0.02, the value of 1.58 eV was observed for both grain and grain boundaries, which explains the convolution of the Nyquist plot for this composition (Fig. 8(b)). Moreover, the nearly same values of the activation energies for the resistance and the relaxation time for the same composition reveal that the conduction and the relaxation time are described by the same mechanism.

(Insert Fig. 10, here)

(Insert Table 5, here)

## 3.5 Electric conductivity

The ac conductivity displays similar frequency dependence in the measured temperature range from 300°C to 360°C of all BSTS samples (Fig. 11(a-d)). Three distinct regions can be observed in the conductivity spectrum, (i) at low frequencies, the ac-conductivity shows a plateau which is independent of frequency ($\sigma_{dc}$) (region I), (ii) in the middle region (region II), the conductivity increases non-linearly with the frequency and (iii) at higher frequencies (region III) the conductivity exhibits high frequency dispersion. For the studied samples, the electrical conductivity increases with frequency and temperature.

Insets to Figs. 10(a-d) show the variation of $\sigma_{dc}$ vs. 1/T of BSTS ceramics, which follow the Arrhenius equation

$$\sigma_{dc} = \sigma_0 \exp(-E_a/k_B T), \qquad (6)$$

where $\sigma_0$, $k_B$ and $E_a$ are the pre-exponential factor, the Boltzmann constant and the activation energy, respectively. A linear behavior of $\sigma_{dc}$ vs. 1/T is observed for the studied samples. The values of the activation energies are in the same range as for grain resistances and relaxations.

(Insert Fig. 11, here)

## 4. Conclusions

The polycrystalline $Ba_{0.9}Sr_{0.1}Ti_{1-x}Sn_xO_3$ (x = 0, 0.02, 0.05 and 0.10) ceramics have been prepared through semi-wet route at relatively low temperature. The analysis of XRD patterns by Rietveld refinement reveals that BSTS samples exhibit T (P4mm) phase for both compositions (x = 0 and 0.02), coexistence of T (P4mm) - O (Amm2) phases for x=0.05 and C (Pm-3m) phase for x=0.10 at room temperature. SEM micrographs illustrate an increase in grain size for 0≤x≤0.05 with clear grain boundaries in contrast to the composition of x=0.10, which exhibits inhomogeneous grains. These SEM results are in good agreement with dielectric measurements, which show an increase in the maximum value of the dielectric permittivity when increasing Sn-content, followed by a decrease and broadening of the dielectric peak for x = 0.10. Direct electrocaloric measurements were performed on three studied samples (x = 0, 0.02 and 0.05), which show an increase in EC response with increasing Sn rate. The x = 0.05 composition exhibits the highest value of the EC responsivity ($\xi$ = 0.027 K cm/kV) under relatively low applied electric field of 7 kV/cm. This value is comparable to that observed in lead-based perovskites such as PMN-PT or PLZT demonstrating the promising potential of BSTS in electrocaloric cooling applications. Impedance spectroscopy and ac conductivity measurements show that the conduction and relaxation processes can be described by the same activated mechanism.

## Acknowledgements

The authors gratefully acknowledge the generous financial support of the European H2020-MSCA-RISE-2017-ENGIMA, ARRS project J1-9147 and program P1-0125 and the CNRST Priority Program (PPR15/2015).

**Table captions**

Table 1: Lattice parameters and the unit cell volume of BSTS samples.

Table 2: Ceramic bulk density, theoretical density and relative density of BSTS samples.

Table 3: The fitted parameters of BSTS ceramics at different compositions $x$ measured at 1 KHz.

Table 4: ECE characteristic parameters obtained in this work in comparison with literature.

Table 5: The activation energies calculated from fitting parameters of the Arrhenius law.

**Figure captions**

Figure 1: (a) XRD patterns and the corresponding Rietveld refinement of BSTS samples calcined at 950°C, (b) Enlarged view of the peak near $2\theta \approx 45°$ for the studied samples.

Figure 2: The evolution of lattice parameters and the unit cell volume with Sn-content (x=0-0.10).

Figure 3: SEM micrographs of (a) BSTS0, (b) BSTS2, (c) BSTS5, and (d) BSTS10 ceramics sintered at 1350°C.

Figure 4: Temperature dependence of the dielectric permittivity and dielectric loss for BSTS ceramics: (a) BSTS0, (b) BSTS2, (c) BSTS5, and (d) BSTS10.

Figure 5: Theoretical and experimental curves of ε' vs. T for all the compositions $x$. Inset shows the evolution of the room temperature (RT) permittivity as a function of the composition $x$.

Figure 6: Temperature dependence of 1/ε' for all BSTS samples.

Figure 7: Electrocaloric temperature change $\Delta T_{EC}$ versus temperature at applied electric field of 7kV/cm for BSTS0, BSTS2 and BSTS5.

Figure 8: Nyquist plots of BSTS samples: (a) BSTS0, (b) BSTS2, (c) BSTS5, and (d) BSTS10 at different temperatures. The fitted curve is shown in the insets for a representative temperature (360 °C).

Figure 9: Variation of the resistance of grain as a function of the inverse of temperature for BSTS samples. Inset depicts the resistance of grain boundaries vs. 1/T of BSTS samples.

Figure 10: Variation of the relaxation time of grain vs. 1/T for different BSTS compositions. Inset shows the relaxation time of grain boundaries vs. 1/T for BSTS samples.

Figure 11: Frequency dependence of the ac conductivity of (a) BSTS0, (b) BSTS2, BSTS5, and (d) BSTS10. Insets show the variation of dc conductivity as a function of the inverse of temperature of BSTS samples.

Table 1

| x | Structure | Space group | Unit Cell parameters | | | |
|---|---|---|---|---|---|---|
| | | | a(Å) | b(Å) | c(Å) | V(Å$^3$) |
| 0 | Tetragonal | P4mm | 3.9896 | - | 4.0105 | 63.838 |
| 0.02 | Tetragonal | P4mm | 3.9943 | - | 4.0109 | 63.994 |
| 0.05 | Tetragonal + Orthorhombic | P4mm + Amm2 | 4.0030 3.9786 | - 4.0389 | 4.0123 3.9843 | 64.295 64.027 |
| 0.10 | Cubic | Pm-3m | 4.0079 | - | - | 64.382 |

Table 2

| Composition x | $\rho_b (g/cm^3)$ | $\rho_{th} (g/cm^3)$ | $\rho_r$ (%) |
|---|---|---|---|
| **0** | 5.54 | 5.94 | 93 |
| **0.02** | 5.55 | 5.95 | 93 |
| **0.05** | 5.75 | 5.98 | 96 |
| **0.10** | 5.71 | 6.06 | 94 |

Table 3

| x | $\varepsilon'_m$ | $T_m$ (K) | $T_0$ (K) | Δ (K) | γ |
|---|---|---|---|---|---|
| 0 | 7660.01 | 362 | 351 | 288.88 | 1.13 |
| 0.02 | 9339.98 | 349 | 339 | 288.66 | 1.23 |
| 0.05 | 13260.53 | 326 | 322 | 288.37 | 1.34 |
| 0.10 | 7541.96 | 294 | 293 | 301.48 | 1.70 |

Table 4

| Material | T (K) | $\Delta T_{EC,max}$ (K) | $\Delta E$ (kV/cm) | $\xi$ (K cm/kV) | Refs | method | Synthesis method |
|---|---|---|---|---|---|---|---|
| $Ba_{0.9}Sr_{0.1}TiO_3$ | 380 | 0.1415 | 7 | 0.020 | This work | Direct | Semi-wet |
| $Ba_{0.9}Sr_{0.1}Ti_{0.98}Sn_{0.02}O_3$ | 370 | 0.1024 | 7 | 0.015 | This work | Direct | Semi-wet |
| $Ba_{0.9}Sr_{0.1}Ti_{0.95}Sn_{0.05}O_3$ | 340 | 0.1886 | 7 | 0.027 | This work | Direct | Semi-wet |
| $Ba_{0.92}Sr_{0.08}Ti_{0.9}Sn_{0.1}O_3$ | 303.7 | 0.34 | 15 | 0.022 | [37] | Indirect | Solid-state |
| $BaSn_{0.1}Ti_{0.9}O_3$ | 335 | 0.207 | 10 | 0.0207 | [38] | Indirect | Solid-state |
| $BaSn_{0.12}Ti_{0.88}O_3$ | 315 | 0.27 | 10 | 0.027 | [38] | Indirect | Solid-state |
| $BaSn_{0.15}Ti_{0.85}O_3$ | 290 | 0.237 | 10 | 0.0237 | [38] | Indirect | Solid-state |
| $BaSn_{0.18}Ti_{0.82}O_3$ | 273 | 0.187 | 10 | 0.0187 | [38] | Indirect | Solid-state |
| $Ba_{0.75}Sr_{0.25}TiO_3$ | 320 | 0.15 | 10 | 0.015 | [39] | Direct | Solid-state |
| $Ba_{0.7}Sr_{0.3}TiO_3$ | 306 | 0.16 | 10 | 0.016 | [39] | Direct | Solid-state |
| $Ba_{0.65}Sr_{0.35}TiO_3$ | 288 | 0.18 | 10 | 0.018 | [39] | Direct | Solid-state |

Table 5

| Composition | $E_a^{RG}$(eV) | $E_a^{RGB}$(eV) | $E_{relax}^{G}$(eV) | $E_{relax}^{GB}$(eV) | $E_a^{\sigma_{dc}}$ (eV) |
|---|---|---|---|---|---|
| **0** | 1.13 | 1.03 | 1.13 | 1.24 | 1.02 |
| **0.02** | 1.63 | 1.35 | 1.58 | 1.58 | 1.43 |
| **0.05** | 1.13 | 1.14 | 1.20 | 1.30 | 1.14 |
| **0.10** | 0.52 | 0.63 | 0.60 | 1.11 | 0.60 |

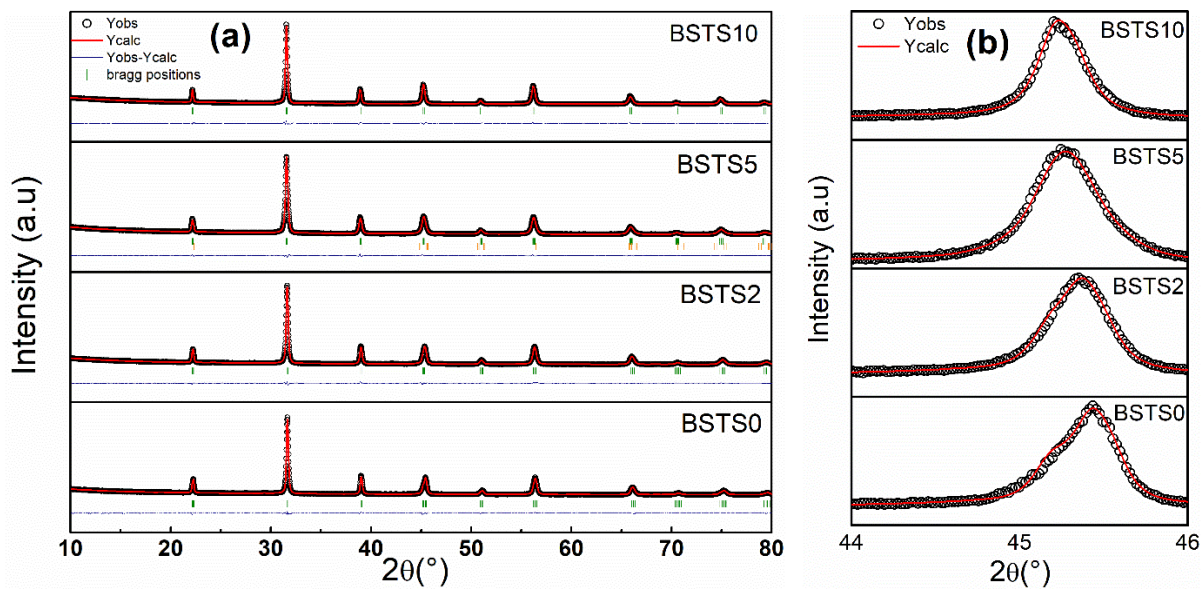

Figure 1

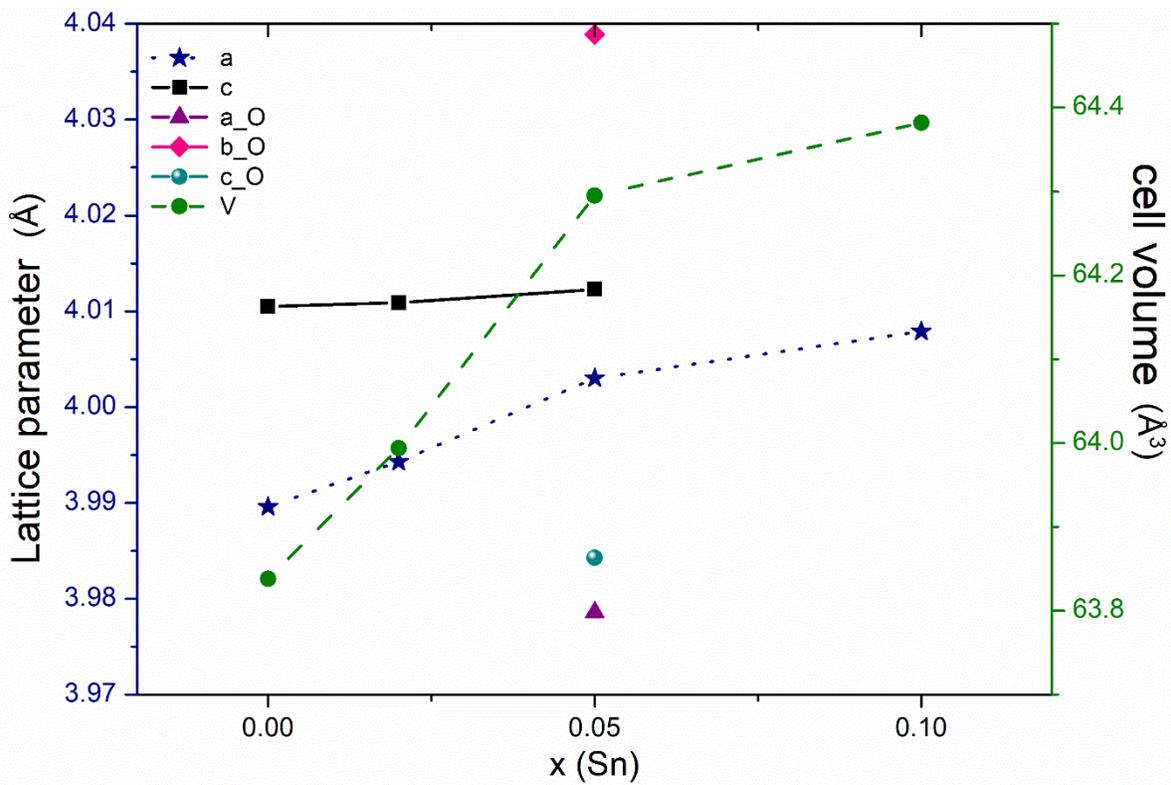

Figure 2

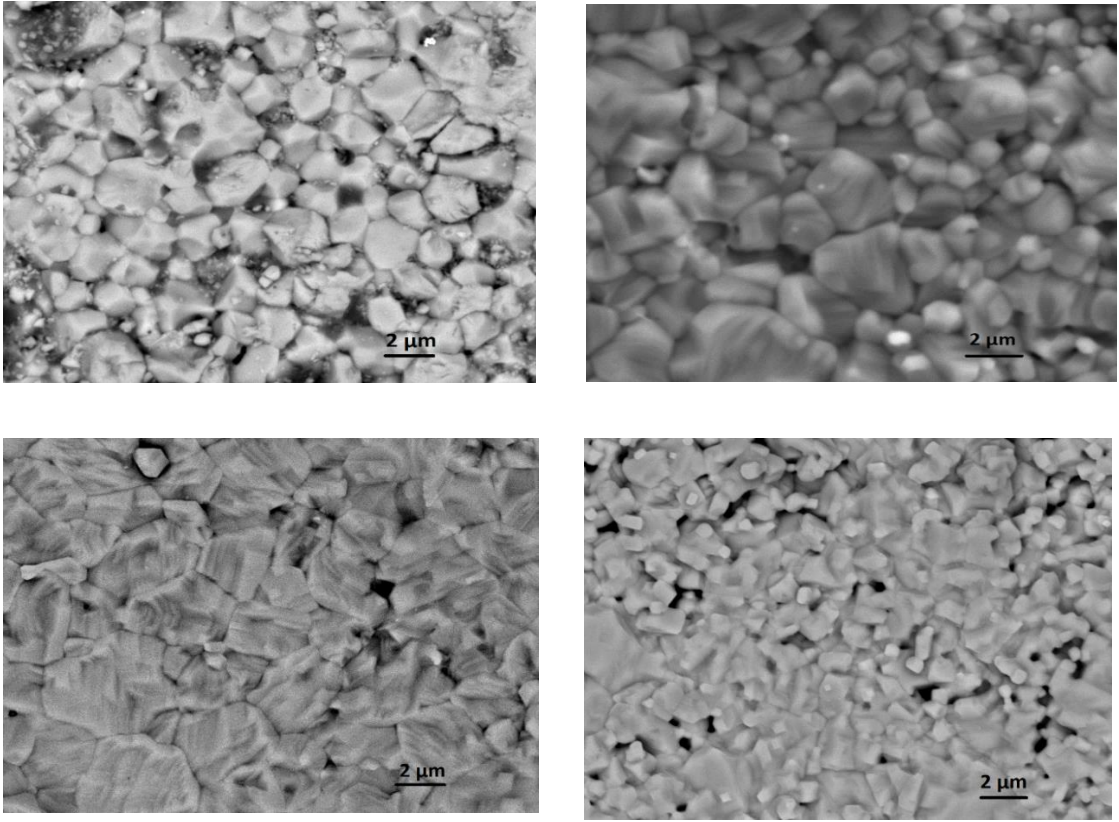

Figure 3
23

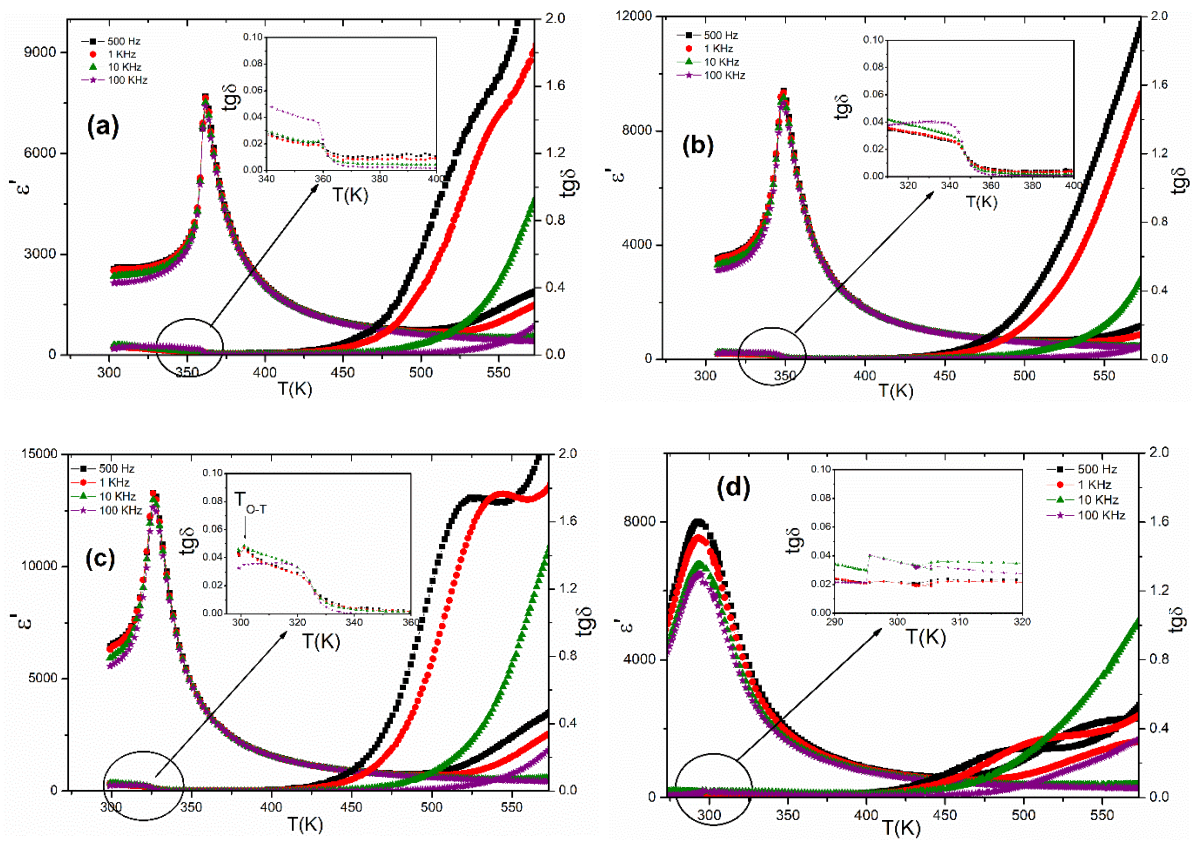

Figure 4

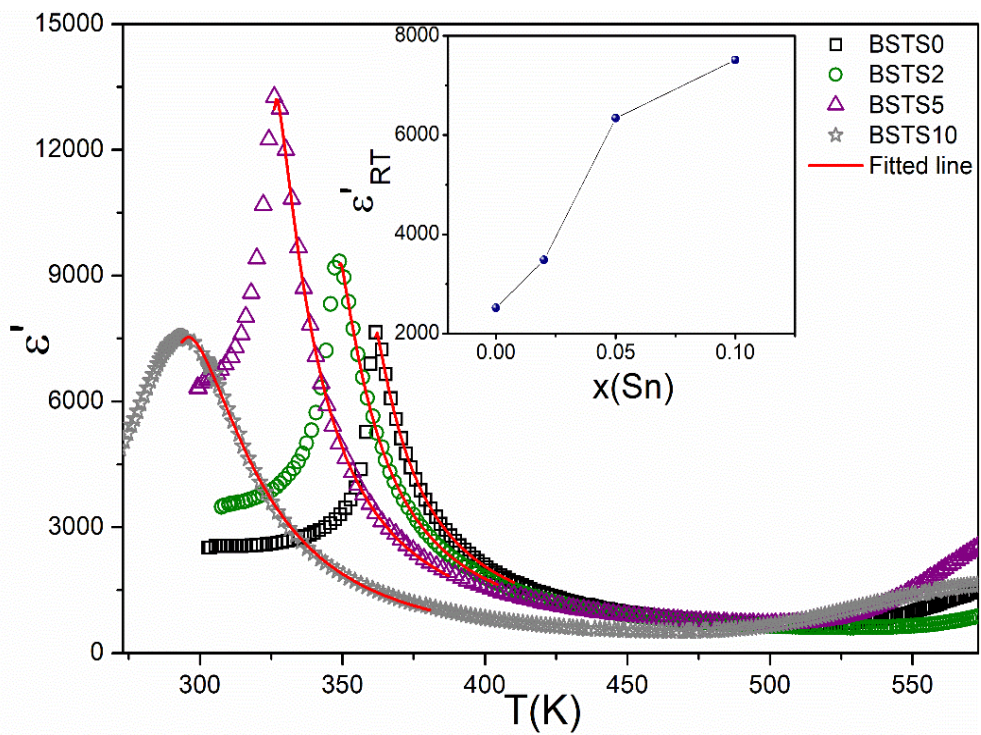

Figure 5

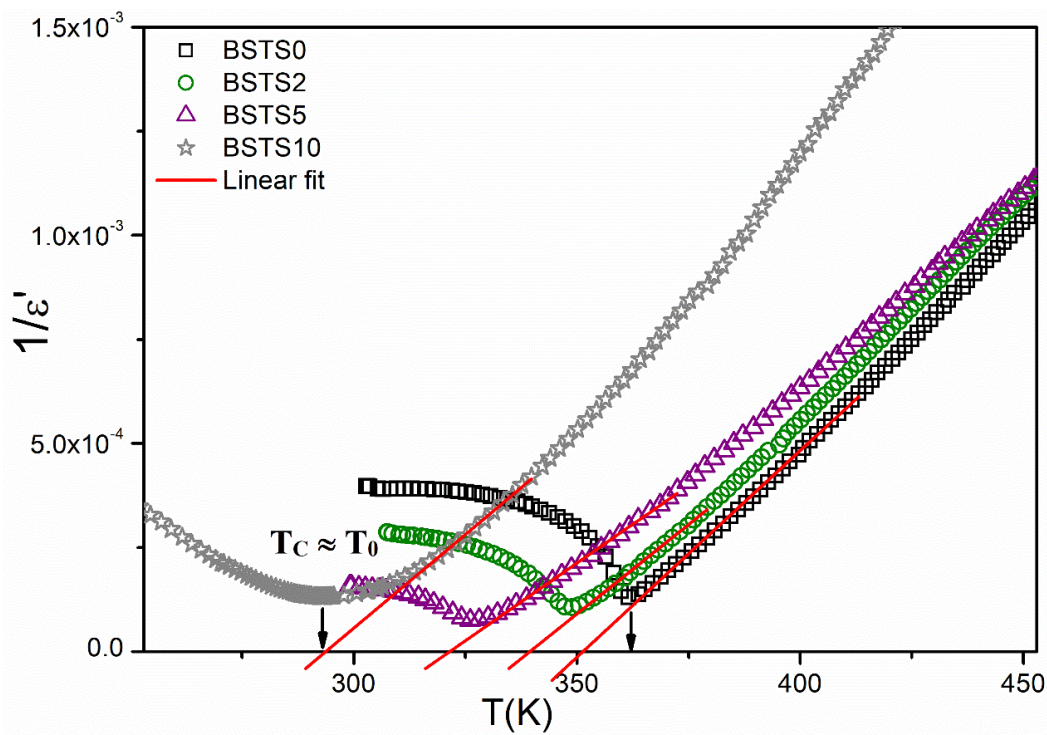

Figure 6

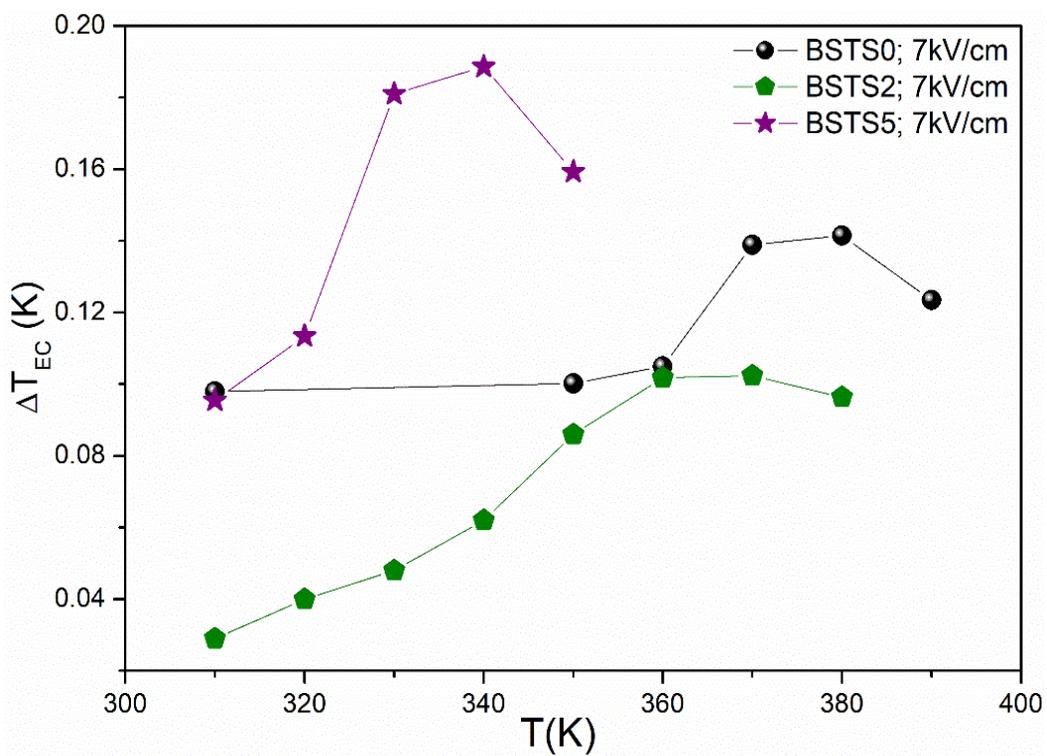

Figure 7

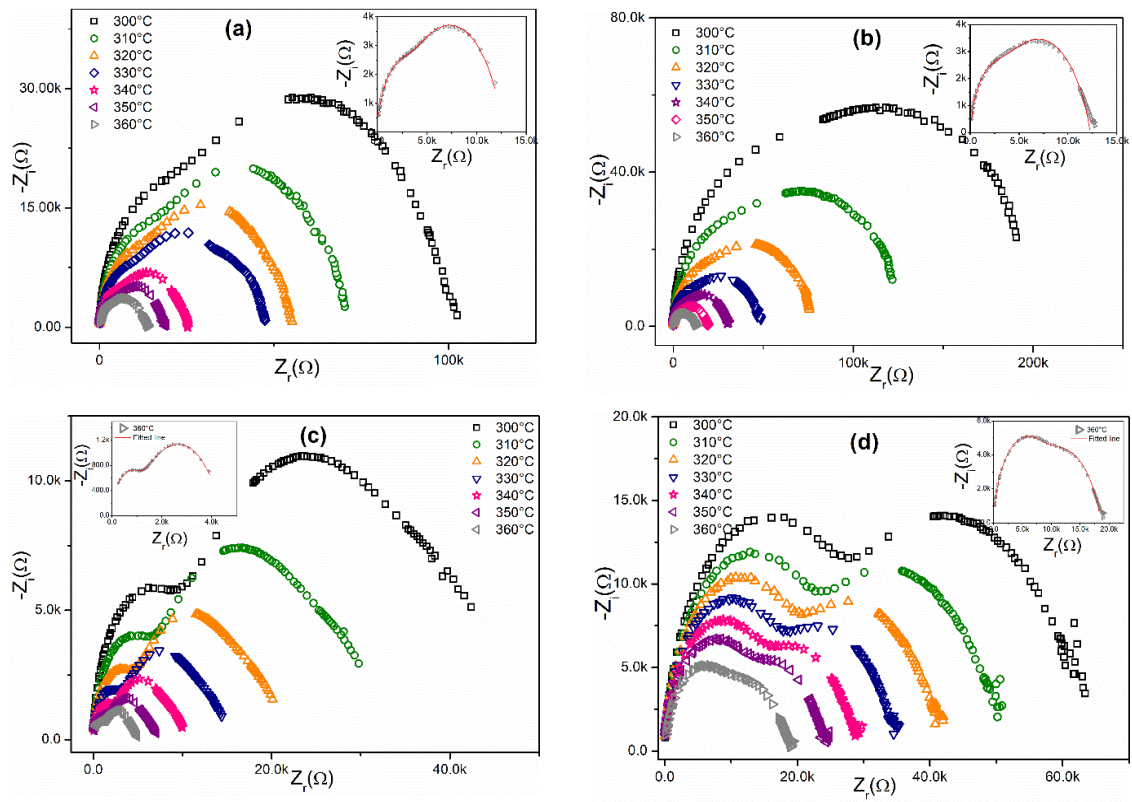

Figure 8

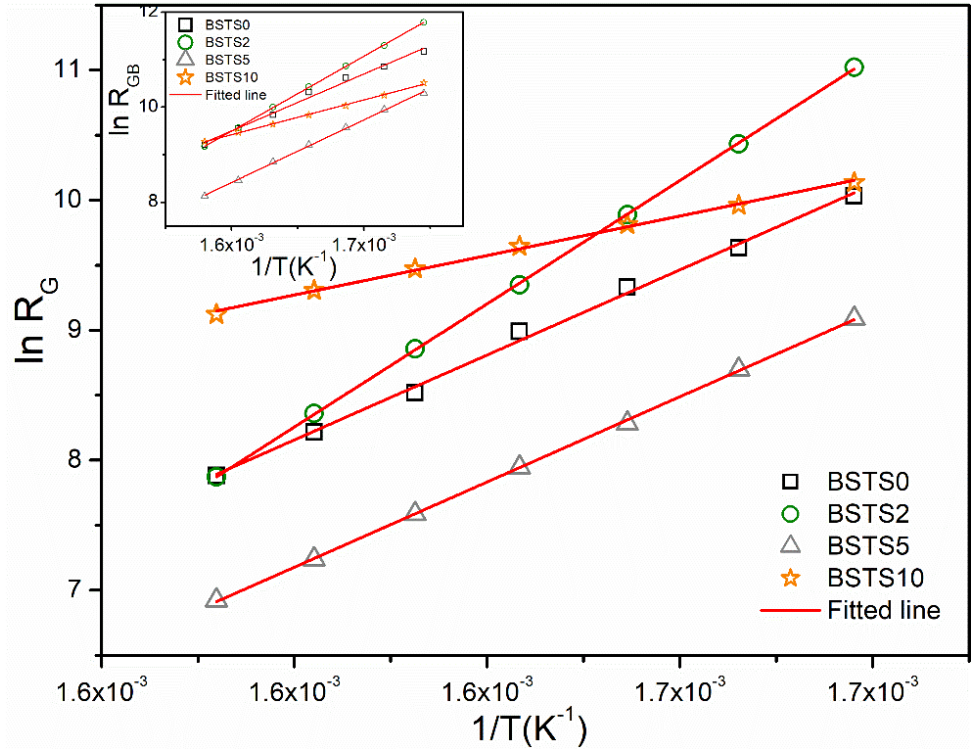

Figure 9

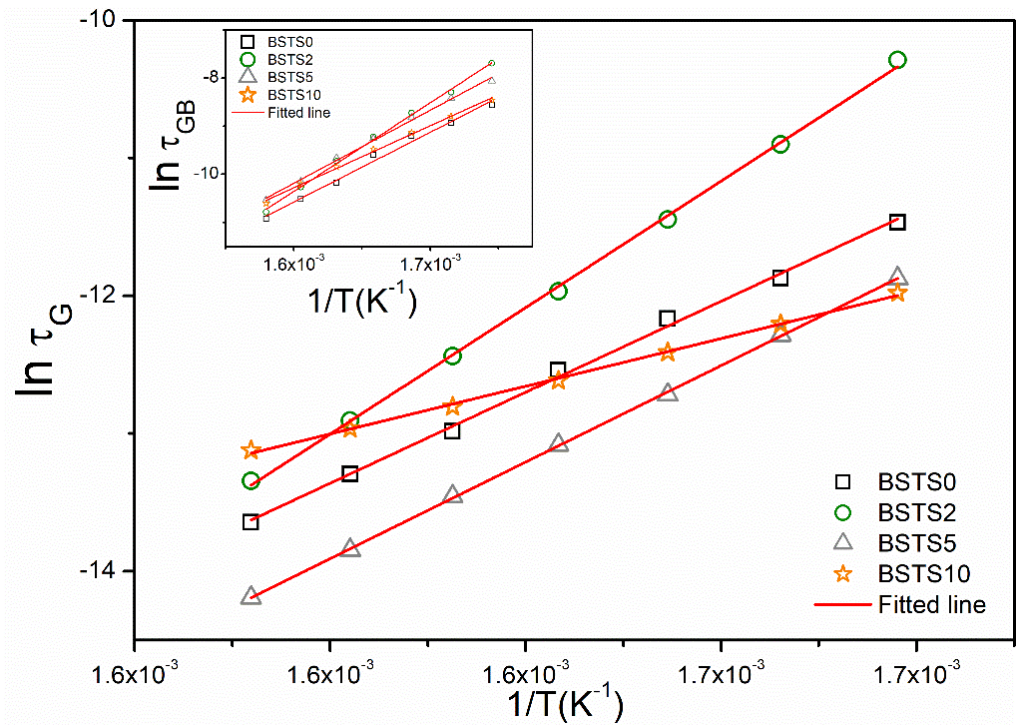

Figure 10

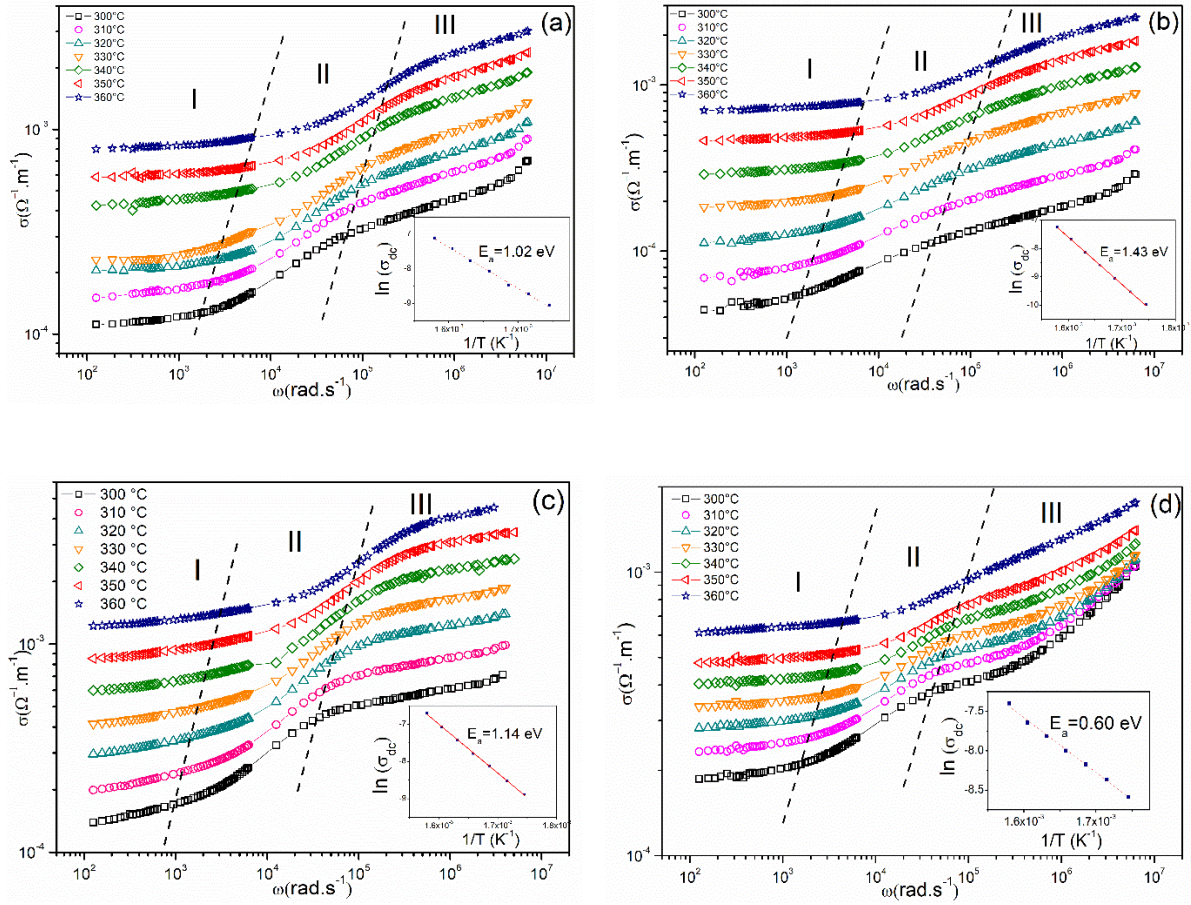

Figure 11